\documentclass[%
 aip,
cp,  
 amsmath,amssymb,
 reprint,%
]{revtex4}

\usepackage{graphicx}
\usepackage{dcolumn}
\usepackage{bm}

\usepackage[utf8]{inputenc}
\usepackage[T1]{fontenc}
\usepackage{mathptmx} 

\usepackage{epsfig}
\usepackage{epstopdf}
\usepackage{subfigure}

\newcommand{\ket}[1]{{\left | {#1} \right\rangle}}
\newcommand{\bra}[1]{{\left\langle {#1} \right |}}

\begin{document}

\title{Benchmarking the Variational Quantum Eigensolver through Simulation of the \\Ground State Energy of Prebiotic Molecules on High-Performance Computers}

\author{P. Lolur} 
 \email{phalgun@chalmers.se}
\author{M. Rahm}%
 \email{martin.rahm@chalmers.se}
\affiliation{
 Department of Chemistry and Chemical Engineering, Chalmers University of Technology, SE-412 96 Gothenburg, Sweden
}

\author{M. Skogh}%
 \email{skoghm@chalmers.se}
\affiliation{Data Science \& Modelling, Pharmaceutical Science, R\&D, AstraZeneca, Gothenburg, Sweden, and \\
 Department of Chemistry and Chemical Engineering, Chalmers University of Technology, SE-412 96 Gothenburg, Sweden
}

\author{L. Garc\'ia-\'Alvarez}
\email{lauraga@chalmers.se}
\affiliation{%
Applied Quantum Physics Laboratory, Department of Microtechnology and Nanoscience-MC2, Chalmers University of Technology, SE-412 96 Gothenburg, Sweden
}%

\author{G. Wendin}
\email [Corresponding author:]{goran.wendin@chalmers.se}
\affiliation{%
Quantum Technology Laboratory, Department of Microtechnology and Nanoscience-MC2, Chalmers University of Technology, SE-412 96 Gothenburg, Sweden
}%

\date{\today} 

\begin{abstract}

We use the Variational Quantum Eigensolver (VQE) as implemented in the Qiskit software package to compute the ground state energy of small molecules derived from water, H$_2$O, and hydrogen cyanide, HCN.
The work aims to benchmark algorithms for calculating the electronic structure and energy surfaces of molecules of relevance to prebiotic chemistry, beginning with water and hydrogen cyanide, and to run them on the available simulated and physical quantum hardware. The numerical calculations of the algorithms for small quantum processors allow us to design more efficient protocols to be run in real hardware, as well as to analyze their performance. Future implementations on accessible quantum processing prototypes will benchmark quantum computers and provide tests of quantum advantage with heuristic quantum algorithms.

\end{abstract}

\maketitle

\section{\label{intro} Introduction}

The project described in this paper is focused on two problems: (1) water (H$_2$O) and water complexes, and (2) hydrogen cyanide (HCN) and HCN-polymerisation. Both topics are of central interest in biochemistry: proton transport in water complexes are of fundamental biological importance, and hydrogen cyanide is a suspected key intermediate in prebiotic chemistry \cite{Sutherland2016,Rahm2016}.

The quantum-computer (QC) quantum chemistry (QChem) efforts are motivated by the significant challenge of accurately calculating electron interactions and correlation energies in several important chemical systems \cite{Wendin2017,Cao2019,McArdle2020}. These systems include studies of bond formation/cleavage, excited state chemistry, reaction pathways, radicals, and ions, among others. Such systems are quantitatively described by the full configuration interaction (Full CI) expansion of the wavefunction, where the wavefunction is an expansion of all the Slater Determinants that can be generated in a given one-electron basis. However, since the number of Slater Determinants depends exponentially on the number of electrons and orbitals, the full CI framework is only effectively utilized for small to medium-sized molecules.
In practice, the active orbital space and the number of correlated electrons are truncated. The largest ever CI calculation that has been reported was a calculation of the pentacene system with an active space of 22 electrons in 22 orbitals, corresponding to a staggering 497634306624 ($\sim 5 \cdot 10^{11}$) Slater Determinants, performed on thousands of processors on a state-of-the-art supercomputer \cite{Vogiatzis2017}. The bottleneck at the core of this challenge is not necessarily just that we are lacking computational hardware but our fundamental inability to express the exponentially increasing description of the wavefunction efficiently.

Calculation of ground-state energies with a QC have been estimated to require 20-50 qubits for H$_2$O and 30-60 qubits for HCN, depending on the basis set and the accuracy needed \cite{Tavernelli2017}. To achieve chemical accuracy (1 kcal/mol, 1.6 mHa, 0.043 eV) in a computation is very challening and requires calculations near the basis set limit. Calculations for H$_2$O with a minimal basis set need only 8 qubits and have been published by other groups (most recently by IBM) \cite{Sokolov2020}, including a number of other small molecules, like nitrogen, N$_2$. Computing the ground state energy of N$_2$ with 12 qubits and a few thousand 2-qubit gates should be possible in the near future with NISQ hardware (HW).

The case of HCN is a bit more challenging for real HW, currently needing 15 qubits and 33 000 2-qubit gates already with a minimal basis set. However, the number of gates and time-to-solution can certainly be reduced with lower demands for accuracy. This may be useful for exploring larger systems and approximate implementations on physical HW. Note that minimal basis sets are of no interest if one wants to challenge modern quantum chemistry. In the case of H$_2$O, the present work demonstrates that a better basis set (Pople 6-31G) requires 20 qubits. For HCN, the 6-31G basis set increases the number of qubits to 33, with 3000 variational parameters and over 600000 gates. This quantum simulation cannot be handled by any high-performance computer (HPC) today. This is remarkable in the light of that an HPC can easily solve the same problem using modern quantum chemistry methods.

\section{Methods}

We have implemented the Python-based Qiskit software package \cite{Qiskit2020,Bravyi2017} on local workstations and clusters, setting up and performing ground-state calculations for water (H$_2$O) and hydrogen cyanide (HCN), as well as for several related molecules and radicals, using the Variational Quantum Eigensolver (VQE) \cite{Peruzzo2014}:
(1) Constructing the Hamiltonian $\hat H$ and a parametrized trial wave function $\ket{\psi(\theta)}$;
(2) Evaluating the energy $E$ of the state $\ket{\psi(\theta)}$, i.e. the expectation value of the Hamiltonian $\hat H$;
(3) Updating the parameters $\theta=(\theta_1,\theta_2,\dots,\theta_m)$ to minimise the energy $E$.
The first and third steps are performed on a classical computer, while the second step is performed on a simulated QC.

The Variational Quantum Eigensolver (VQE) implements the Rayleigh-Ritz variational principle for analysing the energy $E$ for a quatum state $\ket{\psi}$ with respect to the ground state energy $E_0$ of a given Hamiltonian $\hat H$: 

\begin{equation}
E =   \bra{\psi}\hat H \ket{\psi}  \ge  E_0; \;\;\;\;  \hat H = \sum _i\hat H_i 
\label{Evar}
\end{equation}

The VQE is a classical-quantum hybrid algorithm where the trial function $\ket{\psi}$ is created in the qubit register by gate operations. In a fully quantum HW calculation of the expectation value, the energy is estimated via quantum state tomography of each of the Pauli operator products of $\hat H_i$. In quantum simulations on an HPC, the state vector is available classically, and the expectation value of H can be evaluated directly. The VQE scales poorly for large molecules due to repeated measurements/tomography to form the expectation value of the Hamiltonian terms,  $\bra{\psi}\hat H_i \ket{\psi}$.  Nevertheless, the VQE is the common approach for small molecules with present NISQ HW. The phase-estimation algorithm (PEA) scales better, but involves much deeper circuits, and puts much higher demands on the coherence time of the quantum register \cite{Wendin2017,Cao2019}.

The main steps in our VQE calculations are in principle as follows:
We start from a unitary coupled cluster (UCC) ansatz of the quantum state $\ket{\psi}$ with variational parameter  $\ket{\theta}$:
\begin{equation}
\ket{\psi(\theta)} =  \hat U(\theta)  \ket{\psi_{ref}}  =  e^{T(\theta)-T(\theta)^\dagger}  \ket{\psi_{ref}}
\label{CC}
\end{equation}
where $ \ket{\psi_{ref}}$ is, in our approach, the Hartree-Fock (HF) ground state. The ansatz can be expanded: 
\begin{equation}
T(\theta) = T_1 + T_2 + T_3 + .... + T_N 
\label{CCT}
\end{equation}
producing $1,2,3, ...., N$ electron-hole pairs from the N-electron reference state. Explicitly, for $T_1$ and $T_2$: 
\begin{eqnarray}
T_1 =   \sum_{pq} t(\theta)_{pq} \; c^+_p c_q \label{T1}; \;\;\; 
T_2 = \sum_{pqrs} t(\theta)_{pqrs} \; c^+_p c^+_q c_r c_s \label{T2} 
\label{CCT1T2}
\end{eqnarray}
with $c^+_i$ and $c_i$, fermionic creation and annihilation operators, respectively. The series of terms generates in principle all possible configurations for FCI, producing all possible ground and excited state correlations. The terms shown generate single (S) and double (D) excitations and produce the parametrized UCCSD trial-state approximation that we are using. In particular $ t(\theta)_{pq} = \theta_i$ and $t(\theta)_{pqrs} = \theta_j$ for all combinations of the indices $pqrs$. 

\begin{figure}
\includegraphics[width=5cm]{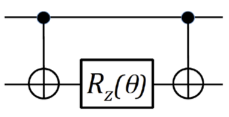} \;\;\;
\includegraphics[width=5.5cm]{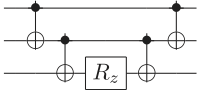}
\caption{Quantum circuits for operator exponentiation:
(a)  $ e^{-i \theta \sigma_{1z} \sigma_{2z}} $; (b) $ e^{-i \theta \sigma_{1z} \sigma_{2z}  \sigma_{3z}} $}
\label{expsigmaz}
\end{figure}

The UCCSD trial-function $\ket{\psi(\theta)}$ with fermionic operators must now be mapped onto qubit spin operators.
Common transformations are the Jordan-Wigner (JW), Bravyi-Kitaev (BK) and Parity encodings, all designed to impose the anticommutation rules. The original UCCSD exponential is then expanded into exponentials of large numbers of products of Paul spin-operators acting on qubits. The parametrized initial trial state is finally constructed through entangled quantum circuits: combinations of parametrized single-qubit rotation gates and entangling CNOT gates (Fig.~\ref{expsigmaz}).  All this results in a state vector $\ket{\psi(\theta)}$ for the trial state.

The fermionic operators $c^+_i$ and $c_i$ in the molecular Hamiltonian
\begin{equation}
\hat H =   \sum_{pq} h_{pq}c^+_p c_q + \frac{1}{2}\sum_{pqrs} h_{pqrs}c^+_p c^+_q c_r c_s 
\label{Atom}
\end{equation}
 must also be expanded in products of Pauli spin-operators using one transformation of the listed above, namely JW, BK and Parity, resulting in the generic interaction form:
\begin{equation}
\hat H =   \sum_{i\alpha} h_{i\alpha}\; \sigma_{i\alpha}   + \sum_{i\alpha,j\beta}  h_{i\alpha,j\beta} \;\sigma_{i\alpha}  \sigma_{j\beta} +  \sum_{i\alpha,j\beta,k\gamma}  h_{i\alpha,j\beta,k\gamma} \;\sigma_{i\alpha}  \sigma_{j\beta} \sigma_{k\gamma}  + .......
\label{Atom_Pauli}
\end{equation}
where $\sigma_{i\alpha}$ corresponds to the Pauli matrix $\sigma_{\alpha}$ for $\alpha \in \{0,x,y,z\}$, acting on the $i$-th qubit.

In practice, we start from a classical HF description and remove states that have the wrong spin and do not conserve the number of electrons (tapering, Z2-symmetry  \cite{Qiskit2020,Bravyi2017}). After fermion-to-spin operator Parity mapping we then Trotterize the UCC-operator (Eq.~\ref{CC}).

The expectation value of the Hamiltonian $ \sum _i\hat H_i $ can then be calculated in two ways: (1) State-vector approach: direct calculation of $ \sum _i \bra{\psi} \hat H_i \ket{\psi}$ by matrix operations (Qiskit state-vector backend); (2) Measurement approach: generating an ensemble of identical trial states and measuring the Pauli operators of the Hamiltonian terms $\hat H_i$ (QASM backend; or experimental q-HW backends).

As a simplest possible example, in the case of the 2-electron hydrogen molecule ($i=1,2)$, one gets: 
\begin{equation}
\hat H =  g_0 {\bf1} + g_1 \sigma_{10} + g_2 \sigma_{20} + g_3 \sigma_{10}  \sigma_{20} + g_4 \sigma_{1x}  \sigma_{2x} + g_5 \sigma_{1y}  \sigma_{2y}
\label{H2BK}
\end{equation}
where $ g_1-g_5$ are coefficients describing the weights of the terms in the transformed Hamiltonian.  The UCCSD approximation creates the trial function via double excitations from the Hartree-Fock (HF) mean-field reference state |01>, building-in electron pair-correlation effects:
\begin{equation}
\ket{\psi(\theta)} = \hat U(\theta)  \ket{\psi_{ref}}     =     e^{-i \theta \sigma_{1x} \sigma_{2y}} \ket{0 1}
\label{}
\end{equation}

The Hartree-Fock reference state |01> is created through a bit flip operation:  $\sigma_{1z}  \ket{0 0} \rightarrow  \ket{0 1}$. The UCCSD-generating quantum circuit is given by a CNOT, a parametrized $ R_z(\theta) $ rotation, and another CNOT, generating the unitary operator in Fig.~\ref{expsigmaz}(a). The desired form in the UCCSD ansatz for  $\ket{\psi(\theta)}$ in Eq.~\ref{CC} is obtained by additional single-qubit rotation gates.
 In general, for systems with more that 2 electrons, the ansatz and the Hamiltonian will involve products with operators involving more than two qubits. A product of 3 operators is shown in Fig.~\ref{expsigmaz}(b) and is generalized to exponents with tensor products of Pauli operators for $n$ qubits, which generates quantum circuits with n-qubit operations.

In the H$_2$-case there is only one single variational parameter, and the optimization of the energy is trivial. For larger molecules, the number of UCCSD variational parameters can be very large (see Tables~\ref{Table1H2O}-\ref{Table2HCN}), and the optimization loop becomes classically intractable.

In summary, our practical approach is as follows:
\begin{itemize}
\item Basic program package: VQE implemented by Qiskit Aqua \cite{Qiskit2020}.
\item Initial/reference state: Hartree-Fock (HF) provided by PySCF.
\item HF wave-functions calculated in general with a Pople minimal orbital basis STO-6G. However, to achieve higher accuracy, in several cases we used 6-31G, and in a few cases 6-31+G* and 6-31++G*. Those basis sets give better accuracy, but also require a much greater number of qubits.
\item Variational ansatz: Trotterized Unitary Coupled Cluster Singles and Doubles (UCCSD). We have chosen to systematically use the UCCSD, rather than experimenting with "hardware-efficient" trial functions \cite{Kandala2017}. UCCSD represents a fundamental QChem benchmark, providing a systematic approximation of many-electron correlations beyond the Hartree-Fock mean-field level. In our view, the UCCSD is an important starting point for developing HW-efficient approaches.
\item Fermion-to-qubit transformation: Parity with qubit tapering \cite{Bravyi2017}.
\item Optimizer: Sequential Least-Squares Linear Quadratic Programming (SLSQP). This is a gradient-based method.
\item QC simulation HW:  iMac i9, 8 cores,128 GB RAM workstation; Chalmers C3SE cluster Vera (Intel Xenon Gold 6130, 32-cores per node, 92GB).
\end{itemize}

\section{Applications, results and discussion}

The objectives of the current project are:
(i) to benchmark QChem to test the limits of quantum software (SW) on real-world problems where a quantum computer (QC) will have an advantage; (ii) to benchmark QC simulations on HPCs against problems in prebiotic chemistry.
In this paper we describe calculations and present results  for:

\begin{enumerate}
\item H$_2$O, OH$^-$, H$_3$O$^+$;
\item HCN, CN$^-$, N$_2$;
\item Water cluster components:   H$_2$O-OH, H$_2$O-HO (bent), H$_2$O-OH$_2$;
\item Water chain proton transfer:  [H$_2$O-H-OH$_2$]$^+$; [OHO]$^-$; OHOHO;
\item Molecules with relevance in prebiotic chemistry:   H$_2$CN$^+$, H$_2$CN$^-$, NHCH$_2$;
\item Tough cases with 25-33 qubits: NCHOH$^-$, NC-CN, HCN (6-31G basis);
\item A few really tough cases (6-31+G* basis) with 69-98 qubits: HCN, C=C=C.
\end{enumerate}

The benchmarking results are presented in two types of tables: 

\begin{itemize}
\item Tables~\ref{Table1H2O} and \ref{Table1HCN} list the molecules in order of number of qubits required. For a given basis set, the molecules are then listed also in the order of number of electrons. The result illustrates the dramatic increase in number of qubits when the basis is expanded, to achieve higher accuracy. 
\item Tables~\ref{Table2H2O} and \ref{Table2HCN} list molecules for which we were able to produce groundstate energies within a reasonable time (weeks) on the iMac i9, 8-core workstation. Note that it is important to distinguish between the precision of a given ground-state energy (how well it is converged), and the accuracy of the result when it comes to comparing with experimental energies. In the present work, the energy numbers are converged to better than $10^{-6}$ Hartree (Ha; 27.2 eV), but the actual energies are not close to experimental chemical accuracy (1.6 mHa). 

\end{itemize}

In Tables~\ref{Table1H2O} and \ref{Table1HCN}, the \#gates column presents the total number of gates after the Qiskit level-1 transpiler. Roughly $70-80\%$ are 2-qubit CNOT gates. Tables~\ref{Table2H2O} and \ref{Table2HCN} also display the very large circuit depths, roughly $90\%$ of the transpiler-optimized gate count. The number of variational parameters is rapidly increasing with the number of qubits, which is particularly noticeable when one expands the basis set for a given molecule.

Tables~\ref{Table1H2O} and \ref{Table1HCN} also present times to solution (TTS) at different stages of running Qiskit.  In order for the SLSQP optimizer to be able to compute a gradient value to produce an energy for the next iteration, it needs to evaluate the energy for every element of the parameter vector. The first TTS-column shows the time in seconds TTS$_{1vp}$ for evaluating the groundstate energy for one single element.
The next column shows the time TTS$_{iter}$ for one iteration, after the SLSQP optimizer has gone through all of the elements in the parameter vector:  TTS$_{iter} \approx$ \#varpar x TTS$_{1vp}$.
The last two columns show the number of energy iterations \#E$_{iter}$ needed to reach $10^{-6}$ Ha precision, and the time TTS$_{conv}$ to converge, in a time  TTS$_{conv} \approx$  \#E$_{iter}$ x TTS$_{iter}$.

The reason for displaying the intermediate times TTS$_{1vp}$ and TTS$_{iter}$ is sheer necessity:  the computation times rapidly become excessive with increasing number of qubits and variational parameters. For the iMac i9 8-cores workstation, it is impractical to produce final converged energy results beyond around 20 qubits. However, the TTS for a final converged groundstate energy can be estimated as TTS$_{conv} \approx$  \#E$_{iter}$ x \#varpar x TTS$_{1vp}$.  The value for \#varpar is known, and \#E$_{iter}$ = 10 is a  typical minimal value for the number of iterations needed.
In this way we have computed TTS$_{1vp}$ for up to 29 qubits. In the particular case of the NCCN molecule with 29 qubits and 515611 gates,  TTS$_{1vp}$ took 132 hours (Table~\ref{Table1HCN}) and the estimated TTS$_{conv} \approx$ 424 years.

Tables~\ref{Table2H2O} and \ref{Table2HCN} list HF and VQE-UCCSD groundstate energies, as well as the difference  E$_{VQE}$ - E$_{HF}$. It should be noted that the numerical results are usually strongly basis dependent: the HF-energy via the PySCF calculation of molecular integrals, and the VQE-energy via the number of qubits, variational parameters and gates. The results will be discussed in some detail below.

\subsection{Water molecule, H$_2$O}

Using Qiskit Chemistry with PySCF and UCCSD, we calculate a converged (tolerance 1 mHa) ground-state energy value for H$_2$O (equilibrium geometry) in a minimal basis (STO-6G) using 8 qubits, 30 variational parameters and 22291 gates (1688 CNOTs) in about 1 minute (Table~\ref{Table1H2O}). This makes it possible to map selected parts of the energy landscape of different geometries. 

It is interesting to note that it turned out to be important to make use of molecular symmetry when placing the atoms in the coordinate system. For instance, if the position of the H$_2$O molecule deviates from a symmetric placement of the H-atoms, the number of qubits increases from 8 to 9, and the number of variational parameters and gates doubles. The reason is that the classical input in the form of molecular integrals doubles. The same is true for other symmetric molecules discussed below.

If one wants better accuracy, to compete with modern QChem on HPCs, one must use much better basis sets than the minimal STO-6G one. However, this dramatically expands the active orbital space: using the modest Pople 6-31G basis set leads to 20 qubits, 408 variational parameters and 73274 gates (56330 CNOTs). This now needs 1 min to evaluate every element in the variational parameter vector (and thus 408 min for every energy value in the optimization loop (Table~\ref{Table1H2O}). This amounts to around 4 days for a single converged result for the ground state energy for one specific geometry. The time to solution (TTS) can of course be reduced using multi-CPU-multi-core HPCs. With 400 cores the calculation can be done in 10-20 min.

Table~\ref{Table2H2O} presents ground state energies for OH$^-$, H$_2$O and  H$_3$O$^+$. Although the HF and VQE-UCCSD energies differ substantially, as expected, the differences in these total energies are relatively similar. 
At the same time, the effect of expanding the basis set for OH- and H$_2$O is very noticeable. The difference between the  HF and VQE-UCCSD energies increase by a factor of 2.5 when going from STO-6G to 6-31G, which better describe the low-electron density regions. This effect can be attributed to both the well known Basis-Set-Superposition Error (BSSE), which artificially stabilizes complexes described by small basis sets, and by a better description of electron correlation using UCCSD.

\subsection{Benchmarking water clusters}

\begin{figure}
\includegraphics[width=5cm]{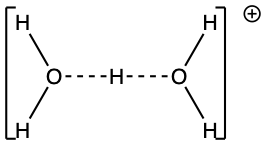}
\caption
{Hydrogen bonding of two water molecules:  [H$_2$O-H-OH$_2$]$^+$.}
\label{H2Ochain}
\end{figure}

Water is a polar molecule and hydrogen bonding between water molecules is essential for understanding the behavior of water and water clusters in biological systems. We have therefore paid special attention to instances involving H-bonding. One such problem is proton transport along water chains via transfer  through hydrogen bonds.

In addition to OH$^-$, H$_2$O and H$_3$O$^+$, we thus computed the groundstate energies of elementary water clusters:   H$_2$O-OH (straight), H$_2$O-HO (bent) and H$_2$O-OH$_2$  (Table~\ref{Table2H2O}). One notes, not surprisingly, that the difference between HF and VQE-UCCSD is slightly larger in the molecular complexes, compared to in the isolated components. The difference is mainly attributable to the BSSE, which artificially lowers the total energy of the molecular complex relative to its constituents.

The next step is to calculate the ground-state energy for [H$_2$O-H-OH$_2$]$^+$  (Fig.~\ref{H2Ochain}), to study proton-transfer along a water chain:
\begin{equation}
H_3O^+ + H_2O \rightarrow  H_2O-H^+-H_2O \rightarrow  H_2O + H_3O^+ 
\label{}
\end{equation}
However, this turned out to be too demanding, limiting us to executing a first energy run for one single  element of the variational parameter vector (Table~\ref{Table1H2O}). This took  3 min in the symmetric case, and 14 min in cases with the central H-atom displaced (H-bonding). Therefore the groundstate energies could not be computed for this molecule.

Instead we considered a few simple model cases, focusing on the central OHO part of [H$_2$O-H-OH$_2$]$^+$  (Fig.~\ref{H2Ochain}), which we take to describe a simplified H-bonding situation (Tables~\ref{Table1H2O} and \ref{Table2H2O}):
\begin{equation}
[O-HO]^- \rightarrow [OHO]^-\rightarrow [OH-O]^-
\label{}
\end{equation}
We find that H-bonded [O-HO]$^-$ has slightly higher energy than symmetric  [OHO]$^-$, meaning that, in our approximation, and with a minimal basis, there is no barrier. The same results are produced at the HF level,  the difference between HF and VQE-UCCSD being very small.  The potential between the O-atoms is very flat, as also demonstrated by the extremely slow convergence of the energy: it took 118 energy iterations to converge for [O-HO]$^-$ and 129 for the symmetric [OHO]$^-$. With a minimal basis set the proton is effectively described as delocalized between the two oxygen nuclei. 
To investigate this more carefully with a larger basis set (e.g. 6-31G) is impossible given the present HPC-resources, as is clear from Table~\ref{Table1H2O}. The number of qubits (32-33) is no problem to handle, but the number of variational parameters is prohibitive. 

We also initiated a calculation of a highly simplified model of concerted proton transfer (Table~\ref{Table1H2O}):
\begin{equation}
OH-OH-O  \rightarrow OHOHO  \rightarrow O-HO-HO 
\label{}
\end{equation}
These molecules need 24 qubits for a minimal basis set (STO-6G) and take about 1 hour per energy evaluation on the 8-core workstation. However, with nearly 700 variational parameters, a single ground-state energy will take 4 weeks. And to converge may take 100 iterations because of the flatness of the energy landscape. Therefore, a super-HPC like SUMMIT will need 3 hours to compute a single point on that approximate energy surface. With a minimal basis set.

\subsection{Benchmarking HCN-isomerization}

\begin{figure}
\includegraphics[width=10cm]{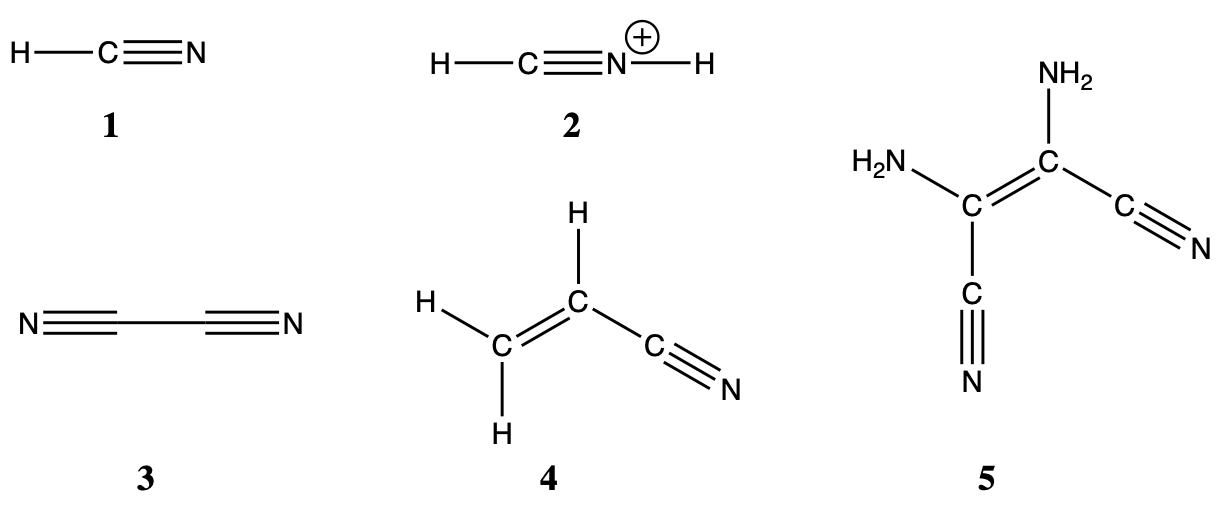}
\caption
{(1) Hydrogen cyanide, HCN; (2) HCNH$^+$; (3) Cyanogene; (4) Acrylonitrile; (5) Diaminomaleonitrile}
\label{HCN+Co}
\end{figure}

A particular task has been to calculate the ground-state energy surface of HCN relevant for HCN isomerization: H-CN $\leftrightarrow$ CN-H. A useful basic comparison involves the 10-electron molecules N$_2$, CN$^-$, HCN, and several HCN isomers described with 12-15 qubits (Table  ~\ref{Table1HCN}).  We have calculated the ground-state energy with the STO-6G basis set at a few H-positions around the CN radical. HNC is computed to lie 1.4 eV above the HCN ground state, with an isomerization barrier (CHN) of 3.7 eV. These values are a factor of 2 larger than the actual ones. The deviation can again be attributed to basis set (STO-6G) deficiency.

We have begun to look at molecules relevant to HCN polymerization reactions (Fig.~\ref{HCN+Co}). In addition to HCN (Fig.~\ref{HCN+Co}(1)) we have calculated the HCNH$^+$ cation (Fig.~\ref{HCN+Co}(2)) and initiated calculations for NCCN (Fig.~\ref{HCN+Co}(3)), as well as for C=C=C, presented in  Table~\ref{Table1HCN}.

We have also considered significantly larger molecules, such as acrylonitrile (Fig.~\ref{HCN+Co}(4)) and diaminomalenonitrile (Fig.~\ref{HCN+Co}(5)). These molecules represent potential examples to be characterized in order to quantify the challenges for HPC QC simulators and quantum HW. For comparison, C=C=C with a fairly good basis set (6-31+G*) needs 98 qubits and 15 million gates (Table ~\ref{Table1HCN}). It should be emphasized that the somewhat  C=C=C molecule is included solely for benchmarking. It is not stable as a free molecule but may form a building block of a carbon wire for connecting atomic-scale components on solid surfaces.

\begin{table}
\caption{H$_2$O and related molecules representing elements of water clusters and water chains: Execution times.
}
\begin{ruledtabular}
\begin{tabular}{ccccccccc}
STO-6G& \# electrons & \# qubits & \#gates  & \#varpar (vp) & TTS$_{1vp}$ (s)  & TTS$_{iter}$  (min) & \# E$_{iter}$ & TTS$_{conv} $ (min)\\
\hline \\
OH$^-$ & 8 & 6 & 492 & 10 & 0.03 & 0.007 &  6 & 0.039  \\
OH$^-$ (6-31G) & 8 &17 & 54290 & 404  & 7.4 & 50 &  10 & 500   \\
H$_2$O & 8 &8 & 2291 & 30 & 0.12 & 0.06 & 9 & 0.52 \\
H$_3$O$^+$   flat & 8 & 10 & 6060 & 66 & -- &  -- & --  & --\\
H$_3$O$^+$  & 8 & 11 & 10900 & 106 & 24 & 42 &  12 & 504 \\
OHO$^-$ & 14 & 14 & 7478 & 68 & 1 & 1.15 & 129 & 148 \\
O-HO$^-$ & 14 & 15 &15089 &  130 & 1.8 & 4 & 118 & 472 \\
H$_2$O-OH$^-$ straight & 16 & 18 & 36112 & 252 & 10 & 42 & 10 & 420 \\
H$_2$O-OH$^-$ bent & 16 & 19 &68162 & 492 & 31 & 254 & 12 & 3048 \\
H$_2$O-OH$_2$ & 16 & 19 & 36382 & 228  & 15 & 57 & 9& 513 \\
H$_2$O (STO-31G) & 8 & 20 & 73274 & 408 & 60 & 408 & 9 & 3937 \\
H$_2$O-HOH & 16 & 20 & 71579 & 460 & 65 & 498 & 10 & 4980 \\
H$_2$O-HOH & 16 & 21 & 111733 & 390 & 390 & 4888 & 10 & 48880 \\
H$_2$OHOH$_2$$^+$ & 16 & 21 & 61184 & 356 & 180 & 1068 & 10 & 10680\\
H$_2$OH-OH$_2$$^+$ & 16 & 22 &116287 & 696 & 840 & 9744 & 10 & 97440 \\
OHOHO & 22 & 24 & 112677 & 678 & 3000 & 33900 & 100 & 3.4 $\cdot$ 10$^6$ \\
O-HO-HO & 22 & 24 &114628 & 678 & 3900 & 44070 & 100 & 4.4 $\cdot$ 10$^6$\\
H$_3$O$^+$ flat (6-31G)  & 8 & 24 & 124787 &  650& 4200 & 45500 & 10 & 0.5 $\cdot$ 10$^6$  \\
H$_2$O (6-31+G*) & 8 & 38 & 503868 & 1726 & -- & -- & -- & --\\
H$_2$O (6-31++G*) & 8 & 44 & 820377 & 2472 & -- & -- & -- & --\\
\end{tabular}
\end{ruledtabular}
\footnotetext[1]{STO-6G is the default basis unless otherwise stated.}
\footnotetext[2]{TTS denotes Time to Solution.}
\footnotetext[3]{TTS$_{iter}\approx$ \#varpar x TTS$_{1vp}$ (if estimated, not computed).}
\footnotetext[4]{TTS$_{conv} \approx$  \#E$_{iter}$ x TTS$_{iter}$ (if estimated, not computed).}
\label{Table1H2O}
\end{table}
\begin{table}
\caption{
H$_2$O and related molecules representing elements of water clusters and water chains: Ground state energies.
}
\begin{ruledtabular}
\begin{tabular}{ccccccccc}
STO-6G& \#electrons & \#qubits & \#gates  & depth & \#varpar (vp) & E$_{HF}$ (Ha)  & E$_{VQE}$ (Ha)&  E$_{VQE}$ - E$_{HF}$ (Ha)  \\
\hline\\
OH$^-$ & 8 & 6 & 492 & 421 & 10 & -74.775318 & -74.798881 &  - 0.023563   \\
OH$^-$ (6-31G) & 8 &17 & 54290 & 48279 & 404 & -75.310965 & -75.440060 &  - 0.129094   \\
H$_2$O & 8 &8 & 2291 & 2003 & 30 & -75.678692 & -75.728533 &  - 0.049841   \\
H$_2$O (6-31G) & 8 & 20 & 73274 & 62076 &  408 & -75.983933 & -76.118828 & - 0.134895   \\
H$_3$O$^+$  & 8 & 11 & 10900 & 9159 &  106 & -76.028821 & -76.088693 &  - 0.059872   \\
OHO$^-$ & 14 & 14 & 7478 & 6521 &  68 & -149.2240724 & -149.383830 & - 0.149282   \\
O-HO$^-$ & 14 & 15 &15089 & 13161 & 130 & -149.225245 & -149.374527 & - 0.143106   \\
H$_2$O$_2$ & 14 & 16 & 24661& 21765 & 181 & -150.168451 & -150.282685 & - 0.114234   \\
H$_2$O-OH$^-$ straight & 16 & 18 & 36112 & 32141 &  252 & -150.382694 & -150.464668 & - 0.081974  \\
H$_2$O-OH$^-$ bent & 16 & 19 &68162 & 60797 &  492 & -150.373919 & -150.455889 & - 0.081970   \\
H$_2$O-OH$_2$ & 16 & 19 & 36382 & 32331 &  228  & -151.097158 & -151.211943 & - 0.114785   \\
\end{tabular}
\end{ruledtabular}
\footnotetext[1]{STO-6G is the default basis unless otherwise stated.}
\footnotetext[2]{H$_2$O, Full CI: E$ \;\approx$ - 76.34 Ha \cite{LiLu2020}.}
\label{Table2H2O}
\end{table}

\begin{table}
\caption{
HCN and related molecules relevant for prebiotic chemistry: Execution times.
}
\begin{ruledtabular}
\begin{tabular}{ccccccccc}
STO-6G& \#electrons & \#qubits & \#gates  & \#varpar (vp) & TTS$_{1vp}$ (s)  & TTS$_{iter}$  (min) & \# E$_{iter}$ & TTS$_{conv} $ (min)\\
\hline \\
N2 & 10 & 12 &11351 & 99 & 1 & 1.8 & 9& 16 \\
CN- & 10 & 13 &17490 & 159 & 1.4 & 4 & 9 & 36 \\
HCN & 10 & 15 & 33407 & 284 & 3 & 14 & 10 & 140 \\
CHN & 10 & 15 & 33447 & 284 & 3 & 14 & 12 & 168\\
C-H-N & 10 & 15 & 33407 & 284 & 3 & 14 & 65 & 910 \\
CNH & 10 & 15 & 36273 & 304 & 3 & 14 & 10 & 140 \\
HCNH$^+$ & 10 & 17 & 59600 & 443 &  7.5 & 55 & 10 & 550 \\
HNCH2 & 12 & 18 &53845 & 376 & 11.5 & 72 & 10 & 720\\
H2NCH2$^+$ & 12 & 20 &78245 & 507 & 67 & 566 & 10 & 5660 \\
C=C=C & 12 & 20 &73169 & 460 & 60 & 460 & 10 & 4600 \\
NCHOH$^-$ & 18 & 25 &233568 & 1427 & 14820 & 352469 & 10 &   3.5 $\cdot$ 10$^6$ \\
N2  (6-31G) & 10 & 28 &223322& 1083 & -- & -- & -- & --\\
NC=CN  & 18 & 29 & 515611&  2815 & 475200 & 22294800 & 10 & 223 $\cdot$ 10$^6$ \\
HCN (6-31G)  & 10 & 33 & 638948 & 2971 & -- & --  &  -- & -- \\
N2  (6-31+G*) & 10 & 64 & 3083796 & 7379& -- & -- & -- & --\\
HCN (6-31+G*)  & 10 & 69 & 6784465 &  16847 & -- & -- & -- & -- \\
C=C=C (6-31+G*)  & 12 & 98 & 15552255 & 25844 & -- & -- & -- & --\\
\end{tabular}
\end{ruledtabular}
\footnotetext[1]{STO-6G is the default basis unless otherwise stated.}
\footnotetext[2]{TTS denotes Time to Solution.}
\footnotetext[3]{TTS$_{iter}\approx$ \#varpar x TTS$_{1vp}$ (if estimated, not computed).}
\footnotetext[4]{TTS$_{conv} \approx$  \#E$_{iter}$ x TTS$_{iter}$ (if estimated, not computed).}
\label{Table1HCN}
\end{table}

\begin{table}
\caption{
HCN and related molecules relevant for prebiotic chemistry: Ground state energies.
}
\begin{ruledtabular}
\begin{tabular}{ccccccccc}
STO-6G& \#electrons & \#qubits & \#gates  & depth & \#varpar (vp) & E$_{HF}$ (Ha)  & E$_{VQE}$ (Ha)&  E$_{VQE}$ - E$_{HF}$ (Ha)  \\
\hline \\
N2 & 10 & 12 &11351 & 9908 & 99 & -108.542471 & -108.699357 & - 0.156886  \\
CN- & 10 & 13 &17490 & 15229 &159 & -91.836839 & -91.973197 & - 0.136358   \\
HCN & 10 & 15 & 33407 & 29648 & 284 & -92.572537 & -92.739900 & - 0.167363   \\
CHN (triangle) & 10 & 15 & 36273 & 32156 & 304 & -92.439876 & -92.579297 &  -0.139421  \\
C-H-N (straight) & 10 & 15 & 33407 & 29648 & 284 & -91.934193 & -92.333651 & - 0.399458  \\
CNH & 10 & 15 & 33447 & 30041 &  284 & -92.541804 & -92.680076 & - 0.138272   \\
\end{tabular}
\end{ruledtabular}
\footnotetext[1]{STO-6G is the default basis unless otherwise stated.}
\label{Table2HCN}
\end{table}

\section{Summary and conclusion}

The goal of this work is to benchmark algorithms for calculating the electronic structure and energy surfaces of molecules of relevance to prebiotic chemistry, beginning with water and hydrogen cyanide, and to run them on the available simulated and physical quantum hardware. In this paper we have specifically benchmarked the VQE as implemented in the Qiskit software package, applying the VQE to computing the ground state energy of water, H$_2$O, hydrogen cyanide, HCN, and a number of related molecules. The energies have been calculated using the statevector backend.
 
 As is clear from Tables~\ref{Table1H2O} - \ref{Table2HCN}, substantial classical computational resources are needed to simulate these systems on classical HPC quantum simulators. 
It is evident that for QChem problems even small numbers of qubits lead to large numbers of gates. And this is then amplified by the variational procedure with many parameters and iterations. The large number of gates will severely limit the types of molecules that can be used for benchmarking real quantum HW. And it will also limit what can be simulated on HPC quantum simulators. QChem problems will provide serious challenges and benchmarks for testing HPC and quantum HW implementations.

The main results of this paper are produced with an iMac i9, 8-cores workstation with 128 GB RAM (max 32 qubits). The practical limit for the iMac to achieve an accurate ground-state energy minimum is so far connected with  H$_2$O (6-31G): 20 qubits and 73 000 gates, taking 66.5 hours to converge to 10$^{-6}$ Ha precision of  the ground-state energy for the equilibrium geometry. 

A few molecules (25-29 qubits) have only been tested for a single state-vector run to determine the time for a single energy evaluation (order of hours days, weeks).  In the toughest case considered, NCCN (STO-6G: 29 qubits, 2815 variational parameters, 515611 gates) (Table~\ref{Table1HCN}), it took 132 CPU-hours (5.5 days) to produce the first energy value. This must then be multiplied by 2815 to evaluate one gradient vector for the first energy estimate by the optimizer. And then this will need to be iterated at least 10 times to converge to an accurate ground state energy value. The time-to-solution (TTS) is then about $3.7 \cdot 10^6$ CPU-hours. For comparison, the world's most powerful HPC, SUMMIT at ORNL, USA, has 9,216 22-core CPUs, in total 202752 cores. Assuming, for simplicity, that those cores can work perfectly in parallel and have the same speed as the iMac i9 CPU (3.6 GHz), then SUMMIT is 202752/8 = 25344 times faster than the iMac and would finish our NCCN-instance in 147 hours = 6 days. In the QChem case, the challenge lies in the number of variational parameters and the number of gates needed.

As a final Qiskit benchmark, we were able to run HCN (6-31+G*) (69q) and C=C=C (6-31+G*) (98q) up to the start of the statevector backend, generating transpiled quantum circuit gate lists in 34 and 229 hours, respectively (Table~\ref{Table1HCN}).

 The general problem is primarily not the required memory size nor the number of qubits but rather
 \begin{itemize}
 \item  the large number of  terms needed to map a fermionic molecular Hamiltonian onto a qubit register,
 \item  the large number of  parameters needed to expand the electronic wavefunction, 
 \item  the large number of energy evaluations and iterations needed to converge to a single point on the energy surface. \\
 \end{itemize}
 
To utilize VQE and achieve near chemical accuracy will be extremely challenging for NISQ processors. 
It is problematic or impossible to achieve chemical accuracy with conventional HPC VQE-simulators already for small molecules such as HCN. But, there is no way around it: one must  benchmark and challenge existing quantum HW and SW with available resources. The water molecule is a kind of "gold standard" even for forefront HPC applications, and H$_2$O is an excellent candidate for testing the VQE on quantum HW. 
Nevertheless, at the present stage of the NISQ era, one has to start with "easy" applications and simple approximations just to benchmark the quantum HW.

For pure benchmarking without chemical relevance, one can use a minimal STO-3G basis set with 8 qubits and a small number of variational parameters and gates. A recent application to an ion trap \cite{Nam2020} demonstrated how the VQE groundstate energy (STO-3G) approached the exact diagonalization result for about 15 configurations (variational parameters). Quantum circuits for the 3 first configurations (determinants) were then implemented on the ion trap quantum HW, with the experimental results agreeing with the classical VQE simulation. 
A slightly more challenging benchmark, with a minimal STO-6G basis set, requires 8 qubits and  30 variational parameters, generating 2300 gates for the trial function within the UCCSD approximation (Table~\ref{Table2H2O}).

The concept of "hardware-efficient" trial functions  \cite{Kandala2017} is an attempt to short-circuit systematic UCCSD approaches and still introduce essential electron correlation.
The recently developed adaptive VQE  \cite{Grimsley2019,Tang2020} provides a more systematic approach to including electron correlation processes in order of monotonically decreasing  weight.
Nevertheless, the electron-correlation problem is computationally hard (NP-hard), so there is no easy way around it. State-of-the-art HPC computation of accurate molecular energies based on the Schr\"odinger equation defines the resources needed, and they are indeed huge  \cite{Vogiatzis2017,LiLu2020}.  

HPC quantum simulators cannot be more efficient than systematic HPC brute-force Full CI calculations. 
Quantum advantage will be possible by definition as soon as a quantum register exceeds the available RAM memory of an HPC. But to profit from that potential quantum advantage, the QPU must be able to run the q-algorithm to solution, and that will involve a very large number of gate operations even for the VQE. So, this is the ultimate challenge of the NISQ era.

\begin{acknowledgments}
This project has been supported by the European Commission through project  820363:  OpenSuperQ;  through the Knut and  Alice Wallenberg (KAW) foundation through the WACQT project, including support from AstraZeneca and Chalmers University of Technology. We are grateful to Igor Sokolov, IBM Zurich, for in-depth discussions and for sharing his Qiskit code with us. This research relied in part on computational resources provided by the Swedish National Infrastructure for Computing (SNIC) at C3SE. 
\end{acknowledgments}

\nocite{*}

\end{document}